\documentclass{article}
\usepackage[utf8]{inputenc}
\pdfoutput=1
\usepackage{multirow}
\usepackage{algorithmic}
\usepackage{amssymb}% http://ctan.org/pkg/amssymb
\usepackage{pifont}% http://ctan.org/pkg/pifont
\usepackage{listings}
\usepackage[normalem]{ulem}
\usepackage[titletoc,title]{appendix}
\usepackage[letterpaper, margin=1in]{geometry}
\usepackage{microtype}

\title{Counterexamples and Proof Loophole for the C/C++ to POWER and ARMv7 Trailing-Sync Compiler Mappings}
\author{Yatin A. Manerkar \and Caroline Trippel \and Daniel Lustig$^*$ \and Michael Pellauer$^*$ \and Margaret Martonosi\\
Princeton University\quad\quad\quad\quad\quad\quad\quad\quad ${}^*$NVIDIA\\
\{manerkar,ctrippel,mrm\}@princeton.edu\quad \{dlustig,mpellauer\}@nvidia.com}
\date{}

\usepackage[numbers]{natbib}
\usepackage{graphicx}

\begin{document}
\maketitle

\begin{abstract}
The C and C++ high-level languages provide programmers with atomic operations for writing high-performance concurrent code. At the assembly language level, C and C++ atomics get mapped down to individual instructions or combinations of instructions by compilers, depending on the ordering guarantees and synchronization instructions provided by the underlying architecture. These compiler mappings must uphold the ordering guarantees provided by C/C++ atomics or the compiled program will not behave according to the C/C++ memory model. In this paper we discuss two counterexamples to the well-known trailing-sync compiler mappings for the Power and ARMv7 architectures that were previously thought to be proven correct. In addition to the counterexamples, we discuss the loophole in the proof of the mappings that allowed the incorrect mappings to be proven correct. We also discuss the current state of compilers and architectures in relation to the bug. 
\end{abstract}

\section{Introduction}

The C and C++ high-level languages provide programmers with atomic memory operations for writing high-performance concurrent code. Different types of atomic memory operations provide different levels of memory ordering guarantees. Stronger memory ordering guarantees usually correlate to lower performance and weaker memory ordering guarantees to higher performance. Users can utilise different types of atomic memory operations depending on the guarantees and performance they require.

At the assembly language level, C and C++ atomics get mapped down to individual instructions or combinations of instructions by compilers, depending on the ordering guarantees and synchronization instructions provided by the underlying architecture. These compiler mappings must uphold the ordering guarantees provided by C/C++ atomics or the compiled program will not behave according to the C/C++ memory model.

In this report we discuss two counterexamples to the well-known \textit{trailing-sync} compiler mappings for the Power and ARMv7 architectures that were supposedly proven correct~\cite{batty:clarifying}. In these counterexamples, a particular execution of the program is forbidden by C/C++ but incorrectly \textit{allowed} by the compiled program. In addition to the counterexamples, we discuss the loophole in the proof of the mappings that allowed the incorrect mappings to be proven correct. We also discuss the current state of compilers and architectures in relation to the bug. 

Section~\ref{sec:background} provides background information on C/C++ atomic memory operations and the relations in the memory model relevant to the counterexamples, as well as the relevant compiler mappings for ARMv7 and Power.
Sections 3 and 4 discuss the two counterexamples for the trailing-sync mapping (variants of the well-known IRIW and RWC litmus tests respectively) at both the C/C++ level and the Power/ARM level. Section 5 discusses the loophole in the proof of the mappings~\cite{batty:clarifying} that allowed the incorrect mappings to be proven correct. Section 6 discusses the current state of compilers and architectures in relation to the bug, and Section 7 concludes.

\section{Background Information}
\label{sec:background}

\subsection{C/C++ atomics and Relevant Memory Model Relations}

C/C++ introduced \texttt{atomic} objects and operations in the C/C++11 standards, along with a new memory model governing the order in which C/C++ threads are allowed to observe each others' memory accesses. The C/C++ memory model is based on the \textit{data-race-free} and \textit{properly-labeled} models of Adve and Hill and Gharachorloo et al. respectively~\cite{weakordering,gharachorloo:release}, which guarantee sequential consistency for programs that do not contain any data races.

Here we provide a brief overview of the portions of C/C++ atomics and the C/C++ memory model that are relevant to our counterexamples. We refer the reader to the large body of work on the C/C++ memory model for further details~\cite{cppconcurrency,cpp14,battythesis,commoncompiler,batty:overhauling}.

Different operations can be conducted on C/C++ atomic objects, such as load, store, and compare-and-exchange operations. Each such operation can be given a memory order, including \texttt{memory\_order\_seq\_cst}, \texttt{memory\_order\_release}, and \texttt{memory\_order\_acquire}, which represent sequentially consistent (SC) operations, release operations, and acquire operations respectively. Different memory orders provide different guarantees with respect to the ordering of the atomic memory access with other accesses in the program. Conceptually, a release operation is a write which ensures that prior accesses are made visible before the release. Likewise, an acquire operation is conceptually a read which ensures that memory operations after the acquire are made visible after the acquire itself. A release corresponds to granting permission to access a set of shared locations, while an acquire is performed to gain access to a set of shared locations ~\cite{adve:tutorial}. Note that an SC read is also an acquire operation and an SC store is also a release operation. SC operations have further constraints on their execution, which are detailed below.

In a given C/C++ execution, the \textit{rf} relation relates a write operation to a read operation which reads the value of that write. The \textit{mo} relation enforces a total order on all write operations to the same address, and all threads must observe writes to a given address in this \textit{mo} order. The \textit{fr} relation relates a given read to the writes that follow the source write of the read in \textit{mo}-order.

The happens-before relation \textit{hb} is the transitive closure of the sequenced-before relation \textit{sb}, which corresponds to program order on an individual C/C++ thread, and the synchronizes-with relation \textit{sw}, which relates release store operations (and stores in their \textit{release sequence}) to acquire read operations that read the release store (or a store in its release sequence). (Release sequences are not necessary for understanding the counterexamples in this paper.)

The total order on sequentially consistent operations \textit{sc} must obey the following constraints~\cite{cpp14,batty:overhauling}:

\begin{itemize}
    \item It must be a total order on SC operations, so any two SC operations \textit{must} be ordered with respect to each other.
    \item It must be consistent with \textit{hb} and \textit{mo} restricted to SC atomics. In other words, it is forbidden for two accesses to be related by \textit{hb}/\textit{mo} in one direction and \textit{sc} in the other direction. These first two conditions are henceforth referred to as the \textit{consistent\_sc\_order} property, using the terminology of Batty et al.~\cite{batty:clarifying}.
    \item SC reads (i.e. reads annotated with \texttt{memory\_order\_seq\_cst}) must either read from the latest SC write before them in the \textit{sc} order, or they must read from a non-SC write that does not happen-before the latest SC write to that location. This condition is henceforth referred to as the \textit{sc\_accesses\_sc\_reads\_restricted} property, again using the terminology of Batty et al.~\cite{batty:clarifying}.
\end{itemize}

Thus, according to the C/C++ memory model, the \textit{sc} order must be consistent with \textit{hb}, and any two SC operations must be ordered with respect to each other by \textit{sc}. Thus, if a \textit{hb} edge exists between two SC atomics, an \textit{sc} edge \textbf{must} also exist between them in the same direction.

\subsection{C/C++ Compiler Mappings to Power and ARMv7}

The Power and ARMv7 architectures are two well-known architectures in use today. They are notable for their weak memory models, which allow a great deal of reordering and can require careful use of dependencies and synchronization instructions in order to ensure desired outcomes. Here we provide a brief overview of the relevant instructions in the Power and ARMv7 memory models. We refer the reader to existing work on the memory models of these architectures for further details~\cite{sarkar2011,alglave:herd}.

In Power, an \texttt{lwsync} is a fence which cumulatively orders all reads and writes prior to the fence before any writes after the fence. An \texttt{lwsync} does not order writes prior to the fence with respect to reads after the fence. A \texttt{sync} in Power is a fence which cumulatively orders all reads and writes prior to the fence before all reads and writes after the fence. The ARMv7 \texttt{dmb ish} fence is analogous to the Power sync.

There are two commonly-accepted compiler mappings from C/C++ to Power and ARMv7: the \textit{leading-sync} mapping and the \textit{trailing-sync} mapping~\cite{batty:clarifying,sewell:mappings}. The relevant portions of these mappings are provided in Tables~\ref{tab:leadingsync} and \ref{tab:trailingsync} for reference. A notable difference between the Power and ARMv7 versions of each mapping is that ARMv7 does not have an equivalent of the Power lightweight \texttt{lwsync} fence. It only has the heavyweight \texttt{dmb ish} fence which provides orderings mostly equivalent to Power's heavyweight \texttt{sync} fence. Thus, the ARMv7 implementations of store releases and trailing-sync SC stores utilise a \texttt{dmb ish} fence where the corresponding Power mappings use an \texttt{lwsync}.

The ``\texttt{cmp; bc; isync}" and ``\texttt{teq; beq; isb}" instruction sequences are known as \texttt{ctrlisync} and \texttt{ctrlisb} respectively. The combination of a conditional branch followed by an \texttt{isync} (on Power) or an \texttt{isb} (on ARMv7) instruction is enough to enforce that all instructions after the \texttt{isync}/\texttt{isb} begin execution after a load which the branch depends on. This initially appears to be enough to implement the orderings required by C/C++ \texttt{memory\_order\_acquire} primitives, but as our counterexamples show, issues can arise when acquires interoperate with SC atomics.

As stated above, SC loads are also acquires and SC stores are also releases. In addition to providing acquire and release semantics respectively, SC loads and stores must also obey the aforementioned constraints on the total \textit{sc} order. Part of these constraints requires ensuring that an SC store followed by an SC load in \textit{sb} appear in that order to all cores. This requires a heavyweight \texttt{sync} or \texttt{dmb ish} fence between the SC store and the SC load on Power and ARMv7. Such a fence can either be incorporated into the mapping before all SC loads (which gives the \textit{leading-sync} mapping) or after all SC stores (which gives the \textit{trailing-sync} mapping). Only the instruction sequences for SC loads and stores change between the leading and trailing-sync mappings.

\begin{table}[h]
    \begin{center}
    \centering
    \footnotesize
\begin{tabular}{| c | c | c |}
  \hline
  C/C++ Atomic Operation    & Power Mapping                     & ARMv7 Mapping \\\hline
  Load Acquire              & \texttt{ld; cmp; bc; isync}          & \texttt{ldr; teq; beq; isb} \\
  Load Seq Cst              & \texttt{sync; ld; cmp; bc; isync}     & \texttt{dmb ish; ldr; teq; beq; isb} \\
  Store Release             & \texttt{lwsync; st}                & \texttt{dmb ish; str} \\
  Store Seq Cst             & \texttt{sync; st}                  & \texttt{dmb ish; str} \\\hline
\end{tabular}
\end{center}
\caption{Leading-sync compiler mapping from certain C/C++11 atomic operations to Power and ARMv7.}
\label{tab:leadingsync}
\end{table}

\begin{table}[h]
    \begin{center}
    \centering
    \footnotesize
\begin{tabular}{| c | c | c |}
  \hline
  C/C++ Atomic Operation    & Power Mapping                     & ARMv7 Mapping \\\hline
  Load Acquire              & \texttt{ld; cmp; bc; isync}          & \texttt{ldr; teq; beq; isb} \\
  Load Seq Cst              & \texttt{ld; sync}                  & \texttt{ldr; dmb ish} \\
  Store Release             & \texttt{lwsync; st}                & \texttt{dmb ish; str} \\
  Store Seq Cst             & \texttt{lwsync; st; sync}          & \texttt{dmb ish; str; dmb ish} \\\hline
\end{tabular}
\end{center}
\caption{Trailing-sync compiler mapping from certain C/C++11 atomic operations to Power and ARMv7.}
\label{tab:trailingsync}
\end{table}

Examining the mappings at a high level, one can notice the following:

\begin{itemize}
    \item The \texttt{lwsync}/\texttt{sync} or \texttt{dmb ish} prior to a release or SC store ensures that accesses before the release or SC store are made visible to other cores before they observe the release.
    \item The \texttt{ctrlisync}/\texttt{ctrlisb} following a load acquire enforces that all accesses after the acquire begin execution after the acquire.
    \item The extra \texttt{sync}/\texttt{dmb ish} before SC loads or after SC stores enforces ordering between SC stores and subsequent SC loads in program order.
\end{itemize}

Both the leading and trailing-sync mappings were supposedly proven correct by Batty et al.~\cite{batty:clarifying}. However, we discovered a loophole in their proof which allowed the incorrect trailing-sync mappings to be proven correct. The loophole is detailed in Section~\ref{sec:loophole}.

The next two sections discuss the counterexamples we discovered for the trailing-sync mapping. These counterexamples were discovered using a framework~\cite{framework:arxiv} capable of exhaustively enumerating common C11 litmus tests with varied combinations of memory orders and comparing their outcomes against those of the equivalent ISA-level litmus tests (obtained by compiling with a given mapping) on a variety of microarchitectural implementations defined using the $\mu$spec language (as seen in the COATCheck paper~\cite{coatcheck}). Specifically, these counterexamples were observed during runs of the framework on a microarchitecture with Power/ARMv7-like features and using a trailing-sync compiler mapping. The runtimes of the framework are very reasonable.

\section{The IRIW Counterexample}
\label{sec:iriw_cex}

\begin{figure}[t]
\begin{center}
\centering
\small
\setlength\tabcolsep{4pt} % default value: 6pt
    \begin{tabular}{| c  c  c  c |}
    \hline
    \multicolumn{4}{|c|}{\textbf{Initial conditions}: a: x=0, b: y=0}\\ \hline
    T0            & T1            & T2            & T3    \\ \hline
    c: st(x, 1, seq\_cst) & d: st(y, 1, seq\_cst) & e: r1 = ld(x, acquire) & g: r3 = ld(y, acquire) \\
                  &               & f: r2 = ld(y, seq\_cst) & h: r4 = ld(x, seq\_cst) \\ \hline
    \multicolumn{4}{|c|}{\textbf{Outcome forbidden by C/C++}: r1=1, r2=0, r3=1, r4=0}     \\
    \hline
    \end{tabular}
\end{center}
\vspace{-10pt}
\caption{The IRIW counterexample, specifically the case where both of the first loads on the reading cores are acquires. In this figure, memory\_order\_\textit{x} is abbreviated to \textit{x} for brevity.}
\label{fig:iriw_cpp}
\end{figure}

\begin{figure}[t]
  \centering
  \includegraphics[width=\textwidth]{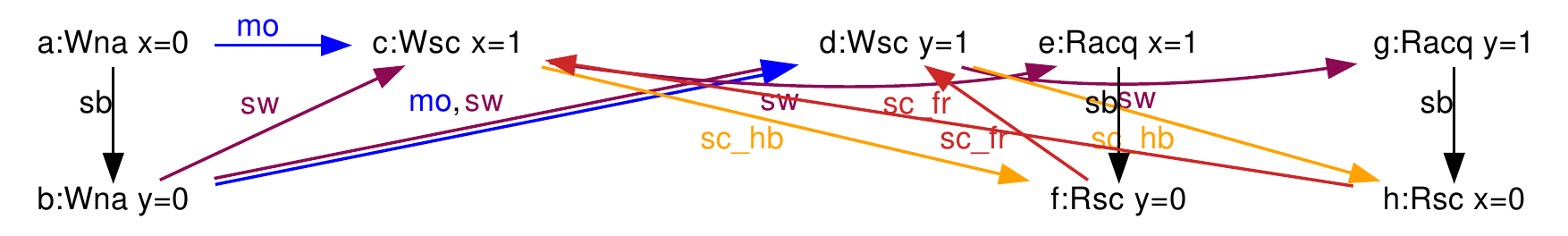}
  \caption{Execution graph of the IRIW counterexample generated with the help of \texttt{CPPMEM}~\cite{Batty:mathematizingc++}, with relevant edges showing why the execution is forbidden by the C/C++ memory model.}
  \label{fig:iriw_graph}
\end{figure}

\begin{figure}[t]
\begin{center}
\centering
\small
\setlength\tabcolsep{3pt} % default value: 6pt
    \begin{tabular}{| c  c  c  c |}
    \hline
    \multicolumn{4}{|c|}{\textbf{Initial conditions}: x=0, y=0}\\ \hline
    C0            & C1            & C2            & C3    \\ \hline
    sync        &  sync & & \\
    \textbf{st x = 1} & \textbf{st y = 1} & \textbf{r1 = ld x} & \textbf{r3 = ld y} \\
                &                   & \textbf{ctrlisync} & \textbf{ctrlisync} \\
                &                   & \textbf{sync} & \textbf{sync} \\
                &                    & \textbf{r2 = ld y} & \textbf{r4 = ld x} \\
                &                   & ctrlisync & ctrlisync \\ \hline
    \multicolumn{4}{|c|}{\textbf{Forbidden}: r1=1, r2=0, r3=1, r4=0}
    \\\hline
    \end{tabular}
    \hspace{20pt}
    \begin{tabular}{| c  c  c  c |}
    \hline
    \multicolumn{4}{|c|}{\textbf{Initial conditions}: x=0, y=0}\\ \hline
    C0            & C1            & C2            & C3    \\ \hline
    dmb ish        &  dmb ish & & \\
    \textbf{st x = 1} & \textbf{st y = 1} & \textbf{r1 = ld x} & \textbf{r3 = ld y} \\
                &                   & \textbf{ctrlisb} & \textbf{ctrlisb} \\
                &                   & \textbf{dmb ish} & \textbf{dmb ish} \\
                &                    & \textbf{r2 = ld y} & \textbf{r4 = ld x} \\
                &                   & ctrlisb & ctrlisb \\ \hline
    \multicolumn{4}{|c|}{\textbf{Forbidden}: r1=1, r2=0, r3=1, r4=0}
    \\\hline
    \end{tabular}
\end{center}
\caption{IRIW counterexample compiled to Power (left) and ARMv7 (right) using the leading-sync compiler mapping. Instructions relevant to the outcome are in bold. The heavyweight \texttt{sync}/\texttt{dmb ish} fences between the pairs of loads on C2 and C3 are sufficient to disallow the forbidden outcome on ARMv7 and Power.}
\label{fig:iriw_leading}
\end{figure}

\begin{figure}[t]
\begin{center}
\centering
\small
\setlength\tabcolsep{3pt} % default value: 6pt
    \begin{tabular}{| c  c  c  c |}
    \hline
    \multicolumn{4}{|c|}{\textbf{Initial conditions}: x=0, y=0}\\ \hline
    C0            & C1            & C2            & C3    \\ \hline
    lwsync        &  lwsync & & \\
    \textbf{st x = 1} & \textbf{st y = 1} & \textbf{r1 = ld x} & \textbf{r3 = ld y} \\
    sync        & sync              & \textbf{ctrlisync} & \textbf{ctrlisync} \\
                &                    & \textbf{r2 = ld y} & \textbf{r4 = ld x} \\
                &                   & sync & sync \\ \hline
    \multicolumn{4}{|c|}{\textbf{Allowed}: r1=1, r2=0, r3=1, r4=0}
    \\\hline
    \end{tabular}
    \hspace{20pt}
    \begin{tabular}{| c  c  c  c |}
    \hline
    \multicolumn{4}{|c|}{\textbf{Initial conditions}: x=0, y=0}\\ \hline
    C0            & C1            & C2            & C3    \\ \hline
    dmb ish        &  dmb ish & & \\
    \textbf{st x = 1} & \textbf{st y = 1} & \textbf{r1 = ld x} & \textbf{r3 = ld y} \\
    dmb ish        & dmb ish              & \textbf{ctrlisb} & \textbf{ctrlisb} \\
                &                    & \textbf{r2 = ld y} & \textbf{r4 = ld x} \\
                &                   & dmb ish & dmb ish \\ \hline
    \multicolumn{4}{|c|}{\textbf{Allowed}: r1=1, r2=0, r3=1, r4=0}
    \\\hline
    \end{tabular}
\end{center}
\caption{IRIW counterexample compiled to Power (left) and ARMv7 (right) using the trailing-sync compiler mapping. Instructions relevant to the outcome are in bold. The absence of heavyweight \texttt{sync}/\texttt{dmb ish} fences between the pairs of loads on C2 and C3 result in the outcome being allowed by both Power and ARMv7 models (and visible on Power hardware).}
\label{fig:iriw_trailing}
\end{figure}

The first counterexample is a variant of the well-known Independent Reads Independent Writes (IRIW) litmus test, where at least one of the first loads on the reading cores is an acquire operation. All other accesses in the test are SC accesses. The case where both of the first loads on the reading cores are acquires is shown in Figure~\ref{fig:iriw_cpp}. The rest of this section focuses on this particular case of the counterexample, though the reasoning for the cases where only one load is an acquire is very similar.

We begin by showing why this outcome is forbidden under the current C/C++ memory model. One execution graph for the test's outcome is shown in Figure~\ref{fig:iriw_graph}. Note that the SC stores c and d synchronize-with the acquire operations e and g respectively, as the acquires read the values of the SC stores, and SC stores are also releases.

As per the current C/C++ memory model, both the \textit{sc\_fr} and \textit{sc\_hb} edges must be part of the \textit{sc} total order. The \textit{sc\_fr} edges (marked in dark red) from $h\rightarrow c$ and $f\rightarrow d$ shadow the \textit{fr} edges between these operations, and must be part of the \textit{sc} order to abide by the \textit{sc\_accesses\_sc\_reads\_restricted} axiom mentioned in Section~\ref{sec:background}. This is because both the SC reads f and h read from the non-SC initial writes (b and a respectively) which \textit{hb} all other writes. If the \textit{sc} order contained the reverse of one of the \textit{sc\_fr} edges, then the initial writes (accesses b and a) that the SC reads read from would \textit{hb} the latest SC writes to the locations (d and c respectively), thus causing the SC reads to fail the \textit{sc\_accesses\_sc\_reads\_restricted} axiom.

Meanwhile, there is a \textit{hb} edge from $c\rightarrow f$ through the transitive composition of the \textit{sw} edge from $c\rightarrow e$ and the \textit{sb} edge from $e\rightarrow f$. Likewise, there is a \textit{hb} edge from $d\rightarrow h$ through the transitive composition of the \textit{sw} edge from $d\rightarrow g$ and the \textit{sb} edge from $g\rightarrow h$. Thus, in order to keep the \textit{sc} order consistent with \textit{hb}, there must also be \textit{sc} edges from $c\rightarrow f$ and $d\rightarrow h$, which correspond to the \textit{sc\_hb} edges in Figure~\ref{fig:iriw_graph}.

The combination of the \textit{sc\_hb} edges with the \textit{sc\_fr} edges results in a cycle in the \textit{sc} order, which means it is not a total order and is thus invalid. As a result, there are no consistent executions of the listed outcome of this C/C++ program as it is impossible to create a total \textit{sc} order for the outcome that abides by both the \textit{consistent\_sc\_order} and \textit{sc\_accesses\_sc\_reads\_restricted} axioms.

The compilation of this test program to Power and ARMv7 using the leading-sync mapping is shown in Figure~\ref{fig:iriw_leading}. The heavyweight \texttt{sync} (or \texttt{dmb ish} in the case of ARMv7) fences between each pair of loads on C2 and C3 are enough to disallow the forbidden outcome of the test in this case. These compiled Power and ARMv7 tests are forbidden by the Power and ARMv7 models of Alglave et al.~\cite{alglave:herd}, and the corresponding outcomes are not observable on Power or ARMv7 hardware.

On the other hand, when the C/C++ program is compiled to Power and ARMv7 using the trailing-sync mapping, the resultant programs are shown in Figure~\ref{fig:iriw_trailing}. In this case, there is only a \texttt{ctrlisync} (or \texttt{ctrlisb} in the case of ARMv7) between each pair of loads on C2 and C3. This is not sufficient to disallow the forbidden outcome of the test, which is allowed by both the Power and ARMv7 models of Alglave et al.~\cite{alglave:herd}. Furthermore, Alglave et al. have observed the forbidden C/C++ outcome of the trailing-sync Power version of this test on Power hardware~\cite{alglave:herd}.

The problem which results in this bug is that the counterexample program induces a \textit{hb} edge between two SC accesses (such as c and f in Figure~\ref{fig:iriw_graph}) by means of the transitive composition of \textit{hb} edges to and from an intermediate non-SC access (in this case, the \textit{sw} and \textit{sb} edges from $c\rightarrow e$ and $e\rightarrow f$ respectively). The requirement of the C/C++ memory model that \textit{sc} be consistent with \textit{hb} thus requires that c be before f in \textit{sc}. In other words, no thread can observe f before it observes c. Both Power and ARMv7 require a heavyweight \texttt{sync}/\texttt{dmb ish} fence between accesses e and f in order to guarantee this property. Similarly, there must exist a \texttt{sync}/\texttt{dmb ish} fence between accesses g and h in order to guarantee the other \textit{sc} edge induced by the requirement that \textit{hb} is consistent with \textit{sc}.

When the test is compiled using the leading-sync mapping, a \texttt{sync}/\texttt{dmb ish} fence is correctly added between each pair of loads to provide the required guarantees. However, in the version of the test compiled using the trailing-sync mapping, there is only a \texttt{ctrlisync}/\texttt{ctrlisb} between each pair of loads, which is not enough to provide the required guarantees. This results in the forbidden C/C++ outcome being allowed by the Power and ARMv7 models (and observable on Power hardware).

\section{The RWC Counterexample}
\label{sec:rwc_cex}

\begin{figure}
\begin{center}
\centering
\small
\setlength\tabcolsep{4pt} % default value: 6pt
    \begin{tabular}{| c  c  c |}
    \hline
    \multicolumn{3}{|c|}{\textbf{Initial conditions}: a: x=0, b: y=0}\\ \hline
    T0            & T1            & T2 \\ \hline
    c: st(x, 1, seq\_cst) & d: r1 = ld(x, acquire) & f: st(y, 1, seq\_cst) \\
                  & e: r2 = ld(y, seq\_cst) & g: r3 = ld(x, seq\_cst) \\ \hline
    \multicolumn{3}{|c|}{\textbf{Outcome forbidden by C/C++}: r1=1, r2=0, r3=0} \\
    \hline
    \end{tabular}
\end{center}
\vspace{-10pt}
\caption{The RWC counterexample. In this figure, memory\_order\_\textit{x} is abbreviated to \textit{x} for brevity.}
\label{fig:rwc_cpp}
\end{figure}

\begin{figure}
  \centering
  \includegraphics[width=\textwidth]{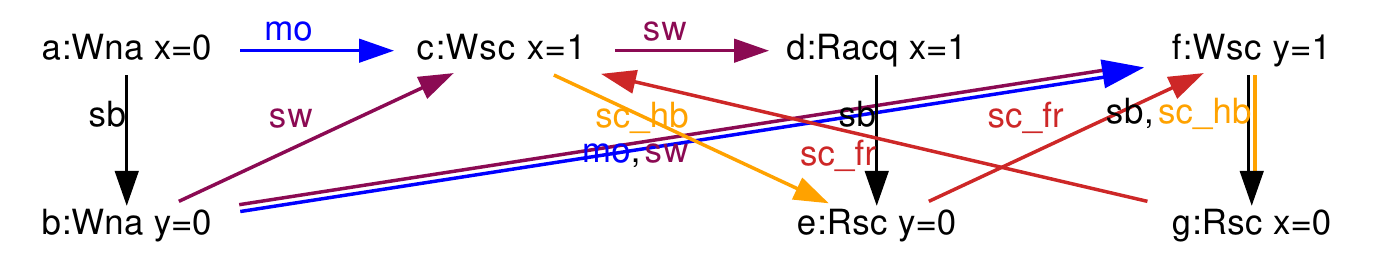}
  \caption{Execution graph of the RWC counterexample generated with the help of \texttt{CPPMEM}~\cite{Batty:mathematizingc++}, with relevant edges showing why the execution is forbidden by the C/C++ memory model.}
  \label{fig:rwc_graph}
\end{figure}

\begin{figure}[t]
\begin{center}
\centering
\small
\setlength\tabcolsep{3pt} % default value: 6pt
    \begin{tabular}{| c  c  c |}
    \hline
    \multicolumn{3}{|c|}{\textbf{Initial conditions}: x=0, y=0}\\ \hline
    C0            & C1           & C2    \\ \hline
    sync        &   & sync \\
    \textbf{st x = 1} & \textbf{r1 = ld x} & \textbf{st y = 1} \\
                & \textbf{ctrlisync} & \\
                & \textbf{sync} & \textbf{sync} \\
                & \textbf{r2 = ld y} & \textbf{r3 = ld x} \\
                & ctrlisync & ctrlisync \\ \hline
    \multicolumn{3}{|c|}{\textbf{Forbidden}: r1=1, r2=0, r3=0}
    \\\hline
    \end{tabular}
    \hspace{20pt}
    \begin{tabular}{| c  c  c |}
    \hline
    \multicolumn{3}{|c|}{\textbf{Initial conditions}: x=0, y=0}\\ \hline
    C0            & C1           & C2    \\ \hline
    dmb ish        & & dmb ish \\
    \textbf{st x = 1} & \textbf{r1 = ld x} & \textbf{st y = 1} \\
                & \textbf{ctrlisb} &  \\
                & \textbf{dmb ish} & \textbf{dmb ish} \\
                & \textbf{r2 = ld y} & \textbf{r3 = ld x} \\
                & ctrlisb & ctrlisb \\ \hline
    \multicolumn{3}{|c|}{\textbf{Forbidden}: r1=1, r2=0, r3=0}
    \\\hline
    \end{tabular}
\end{center}
\caption{RWC counterexample compiled to Power (left) and ARMv7 (right) using the leading-sync compiler mapping. Instructions relevant to the outcome are in bold. The heavyweight \texttt{sync}/\texttt{dmb ish} fences between the pairs of loads on C1 and C2 are sufficient to disallow the forbidden outcome on ARMv7 and Power.}
\label{fig:rwc_leading}
\end{figure}

\begin{figure}[t]
\begin{center}
\centering
\small
\setlength\tabcolsep{3pt} % default value: 6pt
    \begin{tabular}{| c  c  c |}
    \hline
    \multicolumn{3}{|c|}{\textbf{Initial conditions}: x=0, y=0}\\ \hline
    C0            & C1           & C2    \\ \hline
    lwsync        & & lwsync \\
    \textbf{st x = 1} & \textbf{r1 = ld x} & \textbf{st y = 1} \\
    sync        & \textbf{ctrlisync} & \textbf{sync} \\
                & \textbf{r2 = ld y} & \textbf{r3 = ld x} \\
                & sync & sync \\ \hline
    \multicolumn{3}{|c|}{\textbf{Forbidden}: r1=1, r2=0, r3=0}
    \\\hline
    \end{tabular}
    \hspace{20pt}
    \begin{tabular}{| c  c  c |}
    \hline
    \multicolumn{3}{|c|}{\textbf{Initial conditions}: x=0, y=0}\\ \hline
    C0            & C1           & C2    \\ \hline
    dmb ish        &  & dmb ish \\
    \textbf{st x = 1} & \textbf{r1 = ld x} & \textbf{st y = 1} \\
    dmb ish     & \textbf{ctrlisb} & \textbf{dmb ish} \\
                & \textbf{r2 = ld y} & \textbf{r3 = ld x} \\
                & dmb ish & dmb ish \\ \hline
    \multicolumn{3}{|c|}{\textbf{Forbidden}: r1=1, r2=0, r3=0}
    \\\hline
    \end{tabular}
\end{center}
\caption{RWC counterexample compiled to Power (left) and ARMv7 (right) using the trailing-sync compiler mapping. Instructions relevant to the outcome are in bold. The absence of heavyweight \texttt{sync}/\texttt{dmb ish} fences between the pairs of loads on C1 results in the outcome being allowed by both Power and ARMv7 models (and visible on Power hardware).}
\label{fig:rwc_trailing}
\end{figure}

The second counterexample is a variant of the well-known Read-to-Write-Causality (RWC) litmus test, where the first load on the second core is an acquire operation. All other accesses in the test are SC accesses. The C/C++ code for this test is shown in Figure~\ref{fig:rwc_cpp}.

Once again, we begin by showing why the execution is forbidden under the C/C++ memory model. An execution graph for the test's outcome is shown in Figure~\ref{fig:rwc_graph}. As in the IRIW counterexample, the total order on SC operations must include the \textit{sc\_fr} edges as well as the \textit{sc\_hb} edges. The \textit{sc\_fr} edges from $e\rightarrow f$ and $g\rightarrow c$ are required to satisfy \textit{sc\_accesses\_sc\_reads\_restricted}, since both the SC reads read from non-SC writes. Meanwhile, the \textit{sc\_hb} edges from $c\rightarrow e$ and $f\rightarrow g$ are required to satisfy \textit{consistent\_sc\_order}, since there are \textit{hb} edges from $c\rightarrow e$ and $f\rightarrow g$.

The \textit{hb} edge from $c\rightarrow e$ arises through the transitive composition of \textit{sw} and \textit{sb} edges from $c\rightarrow d$ and $d\rightarrow e$ respectively. Meanwhile, the accesses f and g are directly related by sequenced-before (and thus \textit{hb}), and not through an intermediate access.

The combination of the \textit{sc\_fr} and \textit{sc\_hb} edges generates a cycle in the \textit{sc} order, which means it is not a total order as C/C++ requires. Thus, there is no consistent execution of this program that generates the listed outcome under the C/C++ memory model, as it is impossible to construct a correct \textit{sc} order for such an execution.

The compiled versions of this program to Power and ARMv7 using the leading-sync and trailing-sync mappings are shown in Figures~\ref{fig:rwc_leading} and \ref{fig:rwc_trailing} respectively. As in the IRIW counterexample, to enforce the \textit{sc} ordering from $c\rightarrow e$, there must be a heavyweight \texttt{sync}/\texttt{dmb ish} between accesses d and e. In the version compiled with the leading-sync mapping, there are \texttt{sync}/\texttt{dmb ish} fences between each pair of reads (including between accesses d and e), and the overall outcome is correctly forbidden by both the Power and ARMv7 models of Alglave et al.~\cite{alglave:herd}. On the other hand, when compiled with the trailing-sync mapping, there is only a \texttt{ctrlisync}/\texttt{ctrlisb} between accesses d and e, which is not enough to enforce the \textit{sc} ordering from $c\rightarrow e$. Thus, the versions of the test compiled with the trailing-sync mapping incorrectly allow the forbidden C/C++ outcome according to the Power and ARMv7 models of Alglave et al. In addition, Alglave et al. have observed the forbidden C/C++ outcome of the trailing-sync Power version on Power hardware~\cite{alglave:herd}.

It is noteworthy that although the \textit{sc} edge between accesses f and g is required because \textit{sc} must be consistent with \textit{hb}, the accesses are \textit{directly} related through \textit{sb}, and not through an intermediate access. Ordering between SC accesses in program order on the same core is guaranteed by the \texttt{sync}/\texttt{dmb ish} fences used to implement SC accesses in \textbf{both} the leading-sync and trailing-sync mappings. As a result, the required \texttt{sync}/\texttt{dmb ish} fence between accesses f and g exists in the versions compiled using either mapping. Similarly, in the cases of the IRIW counterexample where both of the reads on T2 or T3 are SC reads, a \texttt{sync}/\texttt{dmb ish} fence will always exist between them. It is only in the case where the \textit{hb} edge between two SC accesses arises due to an intermediate non-SC access (as in the case from $c\rightarrow e$ in the RWC counterexample) that the choice of mapping affects the correctness of compilation.

\section{Loophole in the Compilation Proof of Batty et al.}
\label{sec:loophole}

The correctness of compilation from C/C++ to Power (and by analogy, to ARMv7) using both the leading-sync and trailing-sync mappings was supposedly proven by Batty et al.~\cite{batty:clarifying}. Given the above counterexamples, there must be a loophole in the proof that allowed the incorrect mappings to be proven correct. We examined the proof and discovered that the authors did not correctly check whether a given mapping enforced the consistency of the \textit{sc} and \textit{hb} relations with respect to each other. This allowed a mapping like the trailing-sync mapping (which does not always ensure that \textit{sc} is consistent with \textit{hb}) to be proven correct.

As part of their proof, the authors state that the \textit{sc} order is an arbitrary linearization of $(po_t^{sc}\cup co_t^{sc} \cup fr_t^{sc} \cup erf_t^{sc})^*$ (which is -- at a high level -- the combination of program order edges and coherence edges directly between SC accesses). Later in the proof, they state that enforcing that the \textit{sc} order is an arbitrary linearization of the above relation is enough to ensure that \textit{sc} is consistent with \textit{hb}. These claims are not always true. $(po_t^{sc}\cup co_t^{sc} \cup fr_t^{sc} \cup erf_t^{sc})^*$ does not take into account \textit{hb} edges between SC accesses that can arise through the transitive composition of \textit{hb} edges to and from an intermediate non-SC access. (Such edges arise in both our counterexamples.) Per C/C++ memory model requirements, the \textit{sc} order must be consistent with these \textit{hb} edges, but an arbitrary linearization of $(po_t^{sc}\cup co_t^{sc} \cup fr_t^{sc} \cup erf_t^{sc})^*$ may \textit{not} be consistent with them. This also means that enforcing that the \textit{sc} order is an arbitrary linearization of $(po_t^{sc}\cup co_t^{sc} \cup fr_t^{sc} \cup erf_t^{sc})^*$ is \textbf{not} enough to guarantee that the compiled code enforces the constraint that \textit{sc} and \textit{hb} are consistent with each other. This loophole allows the trailing-sync mapping, which does not provide this guarantee (as seen in Sections~\ref{sec:iriw_cex} and \ref{sec:rwc_cex}), to be proven correct.

We contacted Batty et al. regarding this loophole in their proof, and they graciously confirmed our findings.

\section{Current State of Compilers and Architectures}

The previous sections have established that the trailing-sync compiler mapping is invalid for the current C/C++ memory model. Luckily, as of this paper, neither GCC nor Clang implement the exact trailing-sync mapping in Table~\ref{tab:trailingsync} for either Power or ARMv7. Specifically, GCC and Clang use the leading-sync compiler mapping for Power, and while they use a trailing-sync mapping for ARMv7, the mapping for load acquire operations is \texttt{ld; dmb ish} (or stronger). Thus, when compiled for ARMv7 with GCC or Clang, the trailing-sync counterexamples \textbf{do} have \texttt{sync}/\texttt{dmb ish} fences between each pair of loads, which, as outlined in Sections~\ref{sec:iriw_cex} and \ref{sec:rwc_cex}, is enough to disallow the forbidden C/C++ outcome.

Architecturally speaking, the forbidden C/C++ outcomes of both counterexamples have been observed on Power hardware when the tests are compiled using the trailing-sync mapping~\cite{alglave:herd}. While the ARMv7 model of Alglave et al. allows the forbidden C/C++ outcome when the test is compiled to ARMv7 using the trailing-sync mapping, the behaviour has not been observed on ARMv7 hardware~\cite{alglave:herd}.

Finally, while the leading-sync compiler mapping is not vulnerable to the counterexamples discussed in this paper, Vafeiadis et al. have recently also found a counterexample for the leading-sync compiler mapping~\cite{sarkar:personal}, which they will be publishing separately. The combination of these counterexamples means that it is currently \textbf{impossible} to correctly compile C/C++ to Power or ARMv7 with either mapping, and either the mappings or the C/C++ memory model will need to change for correct compilation to be possible.

\section{Conclusion}

In this paper we have outlined two counterexamples for the trailing-sync compiler mappings from C/C++ to Power and ARMv7, as well as the loophole in a prior proof of correctness of these mappings that allowed them to be proven correct. Looking forward, either the mappings or the C/C++ memory model will need to change in order to ensure that the guarantees of the high-level language memory model are respected by compiled programs running on the Power and ARMv7 architectures.

\section{Acknowledgements}

This work was supported in part by C-FAR, one of the six SRC STARnet Centers, sponsored by MARCO and DARPA.

\bibliographystyle{plain}
\bibliography{references}

\begin{thebibliography}{10}

\bibitem{adve:tutorial}
Sarita Adve and Kourosh Gharachorloo.
\newblock Shared memory consistency models: A tutorial.
\newblock {\em {IEEE} {C}omputer}, 29(12):66--76, 1996.

\bibitem{weakordering}
Sarita~V. Adve and Mark~D. Hill.
\newblock Weak ordering -- a new definition.
\newblock In {\em Proceedings of the 17th Annual International Symposium on
  Computer Architecture}, ISCA '90, pages 2--14, New York, NY, USA, 1990. ACM.

\bibitem{alglave:herd}
Jade Alglave, Luc Maranget, and Michael Tautschnig.
\newblock Herding cats: Modelling, simulation, testing, and data mining for
  weak memory.
\newblock {\em ACM Transactions on Programming Languages and Systems (TOPLAS)},
  36(2):7:1--7:74, July 2014.

\bibitem{battythesis}
Mark Batty.
\newblock {\em The C11 and C++11 Concurrency Model}.
\newblock PhD thesis, University of Cambridge, Cambridge, UK, 2014.

\bibitem{batty:overhauling}
Mark Batty, Alastair~F. Donaldson, and John Wickerson.
\newblock Overhauling {SC} atomics in {C11} and {OpenCL}.
\newblock In {\em 43rd Annual Symposium on Principles of Programming Languages
  (POPL)}, 2016.

\bibitem{batty:clarifying}
Mark Batty, Kayvan Memarian, Scott Owens, Susmit Sarkar, and Peter Sewell.
\newblock Clarifying and compiling {C/C++} concurrency: From {C++11} to
  {POWER}.
\newblock In {\em Proceedings of the 39th Annual ACM SIGPLAN-SIGACT Symposium
  on Principles of Programming Languages}, POPL '12, pages 509--520, New York,
  NY, USA, 2012. ACM.

\bibitem{Batty:mathematizingc++}
Mark Batty, Scott Owens, Susmit Sarkar, Peter Sewell, and Tjark Weber.
\newblock Mathematizing {C++} concurrency.
\newblock In {\em 38th Annual Symposium on Principles of Programming Languages
  (POPL)}, 2011.

\bibitem{cppconcurrency}
Hans-J. Boehm and Sarita~V. Adve.
\newblock Foundations of the {C++} concurrency memory model.
\newblock In {\em 29th Conference on Programming Language Design and
  Implementation (PLDI)}, 2008.

\bibitem{gharachorloo:release}
Kourosh Gharachorloo, Daniel Lenoski, James Laudon, Phillip Gibbons, Anoop
  Gupta, and John Hennessy.
\newblock Memory consistency and event ordering in scalable shared-memory
  multiprocessors.
\newblock {\em 17th International Symposium on Computer Architecture (ISCA)},
  1990.

\bibitem{cpp14}
ISO/IEC.
\newblock {Programming Languages -- C++}, 2014.

\bibitem{coatcheck}
Daniel Lustig, Geet Sethi, Margaret Martonosi, and Abhishek Bhattacharjee.
\newblock {"COATCheck: Verifying Memory Ordering at the Hardware-OS Interface}.
\newblock In {\em Proceedings of the 21st International Conference on
  Architectural Support for Programming Languages and Operating Systems}, 2016.

\bibitem{sarkar2011}
Susmit Sarkar, Peter Sewell, Jade Alglave, Luc Maranget, and Derek Williams.
\newblock Understanding power multiprocessors.
\newblock In {\em Proceedings of the 32nd ACM SIGPLAN Conference on Programming
  Language Design and Implementation}, PLDI '11, pages 175--186, New York, NY,
  USA, 2011. ACM.

\bibitem{sewell:mappings}
Peter Sewell.
\newblock {C/C++11} mappings to processors.
\newblock 2016.

\bibitem{framework:arxiv}
Caroline Trippel, Yatin~A. Manerkar, Daniel Lustig, Michael Pellauer, and
  Margaret Martonosi.
\newblock Exploring the trisection of software, hardware, and {ISA} in memory
  model design.
\newblock {\em CoRR}, abs/1608.07547, 2016.

\bibitem{commoncompiler}
Viktor Vafeiadis, Thibaut Balabonski, Soham Chakraborty, Robin Morisset, and
  Francesco Zappa~Nardelli.
\newblock Common compiler optimisations are invalid in the {C11} memory model
  and what we can do about it.
\newblock In {\em 42nd Symposium on Principles of Programming Languages
  (POPL)}, 2015.

\bibitem{sarkar:personal}
Viktor Vafeiadis and Ori Lahav.
\newblock Personal communication, Sept. 27th, 2016.

\end{thebibliography}
\end{document}